# Parametric amplification in single-walled carbon nanotube nanoelectromechanical resonators

Chung-Chiang Wu and Zhaohui Zhong[a]

*Department of Electrical Engineering and Computer Science, University of Michigan,*

*Ann Arbor, MI 48109, USA.*

ABSTRACT

The low quality factor ($Q$) of single-walled carbon nanotube (SWNT) resonators has limited their sensitivity in sensing application. To this end, we employ the technique of parametric amplification by modulating the spring constant of SWNT resonators at twice the resonant frequency, and achieve 10 times $Q$ enhancement. The highest $Q$ obtained at room temperature is around ~700, which is 3-4 times better than previous $Q$ record reported for doubly-clamped SWNT resonators. Furthermore, efficient parametric amplification is found to only occur in the catenary vibration regime. Our results open up the possibility to employ light-weight and high-$Q$ carbon nanotube resonators in single molecule and atomic mass sensing.

---

[a]Author to whom correspondence should be addressed. Electronic mail: zzhong@umich.edu.

With the progress in lithography and material synthesis, nanoelectromechanical systems (NEMS)[1,2] have attracted great interests over the last decade. They can operate at higher frequency with lower power consumption and are expected to have excellent sensitivities in ultra-small mass[3,4] and force sensing[5]. For application in mass spectrometry, a mass sensitivity below a single Delton has been predicted theoretically[6]. In principle, the amount of detected mass ($\Delta m$) is determined by measuring the shift in resonant frequency[1,3] as $\Delta m = 2 m_{eff} \frac{\Delta \omega}{\omega_0}$, where $m_{eff}$ is the effective mass of the resonator, $\Delta \omega$ is the frequency resolution, and $\omega_0$ is the resonant frequency. The frequency resolution can be roughly given by the quality factor as $\Delta \omega = \omega_0/Q$, even though the sensitivity is typically much better. Therefore, $\Delta m$ can then be simply approximated as $\Delta m = 2 \frac{m_{eff}}{Q}$. For better mass sensitivity, one would prefer a resonator with the lowest mass and the highest possible $Q$. To this end, SWNT resonator stands out with one of the highest Young's modulus and the lightest $m_{eff}$ and is considered as a promising candidate[7] to achieve this ultimate goal. $m_{eff}$ for nanotube ($\sim 10^{-18}$ g) is at least three orders of magnitude less than those of other resonators. Unfortunately, the mass sensitivity of SWNT resonators[8-10] is impeded by a poor $Q$. The low $Q$ may result from several possible dissipation sources. Previous studies showed that as structures shrink down to nanometer size scale, the surface effects resulting from the increasing surface-to-volume ratio ($R$) would limit the $Q$[1]. More importantly, strain in nanostructure will generate local temperature difference, leading to irreversibly heat flow along local temperature gradients and inducing the thermalelastic dissipation[1,11]. As the ultimate 1D nanostructure, SWNT resonator's room temperature $Q$ is limited at several dozens. One solution to improve nanotube resonator's sensitivity is to utilize the concept of parametric amplification for $Q$ enhancement.

Parametric resonance is excited by a time-varying modulation of a system parameter. A common example of parametric resonance is a pendulum, with the length of the cord changing with time. If the length decreases when the pendulum is in the lower position and increases in the upper position, oscillations of the pendulum will build up. The first application of parametric amplification in a mechanical resonance system was demonstrated by Rugar and Grütter[12]. In their work, parametric amplification in a mechanical cantilever was obtained by periodically modulating the spring constant on the basis of gate capacitive coupling. Thereafter, numerous studies based on parametric amplification in MEMS resonators have been conducted by many other schemes, for example, exploiting stress via piezoelectric electromechanical coupling[13] or a Lorentz force[14]. Here, we demonstrate parametric amplification in SWNT resonators by modulating the spring constant through a simple electrostatic gate coupling scheme.

To understand parametric amplification in SWNT resonators, we start from the equation of motion. It has been shown that nonlinear dampings[15] are highly important for SWNT and graphene resonators[16]. Here for simplicity, we drop the nonlinear damping terms and model the SWNT resonator as a classical driven damped harmonic oscillator with a time-varying spring constant. The equation of motion is then expressed as[2,12]

$$m\frac{d^2x}{dt^2} + \frac{m\omega_0}{Q}\frac{dx}{dt} + [k_0 + k_p(t)]x = F(t) \qquad (1)$$

, where $k_0$ is the unperturbed spring constant, and $k_p = \Delta k \sin 2\omega_0 t$ is the modulated spring constant created by pumping signals at $2\omega_0$. The parametric amplification will lead to a vibration amplitude gain ($G$) given by[2,12]

$$G = \left[ \frac{\cos^2 \phi}{(1+V_p/V_t)^2} + \frac{\sin^2 \phi}{(1-V_p/V_t)^2} \right]^{1/2} \quad (2)$$

, where $\phi$ is the phase between the driving and pumping signals, $V_t$ is the threshold voltage determined by the system parameters, and $V_p$ is the pumping voltage for parametric amplification. From equation (2), it is expected that the gain will increase as $V_p$ approaches $V_t$ when an appropriate $\phi$ is chosen.

To experimentally verify the parametric amplification, we fabricated SWNT resonators using the one-step direct transfer technique reported previously[17]. For a typical device, the source (S) and drain (D) electrodes are 2 μm wide, separated by 3 μm, 50 nm Au is used as contact metal, and the distance between nanotube and the bottom gate (G) is 1 μm. Detail parameters of similar SWNT resonators can be found in our previous work[18]. To actuate and detect resonance signals of our SWNT resonators, we employed frequency modulation (FM) mixing technique[19] instead of amplitude modulation (AM) method[20] in our measurement setup, as shown in Figure 1(a). For external FM modulation, a small alternating current (AC) signal from a lock-in amplifier at low frequency (616.3Hz) was sent to the RF signal generator. The FM AC signal was then sent to the source electrode to actuate the resonance and the mixing current from the drain electrode was measured by the lock-in amplifier. To achieve parametric amplification, a AC pumping voltage from second RF signal generator was added to the direct current (DC) gate voltage ($V_g$) to modulate the spring constant of the nanotube at $2\omega_0$. We note that the FM mixing technique was chosen as the detection technique because of its better noise-rejection in comparison to the AM method and the background current is zero (an advantage in detecting resonance)[19]. In addition, since noise is affected by the amplitude variation, higher noise level is expected for AM method.

To demonstrate parametric amplification, our SWNT resonator was measured in a vacuum chamber at pressure below $10^{-4}$ torr and $\delta V_{sd} = 20$ mV was applied to drive the nanotube. The mixing current as a function of driving frequency at different pumping voltages are plotted in Figure 1(b). At $V_p = 0$, we observe a nanotube resonance at $f_0 = 23.1$ MHz with a poor $Q$ of 24±1. As we increase the $V_p$ at $2f_0$ frequency, resonance peak amplitude is significantly enhanced and peak width is reduced, indicating a $Q$ enhancement. The $Q$s are extracted by fitting the experimental data of Figure 1(b) with[19]:

$$I(\omega) = \frac{2\omega(\omega^2 - \omega_0^2 - \frac{\omega_0^2}{Q})(\omega^2 - \omega_0^2 + \frac{\omega_0^2}{Q})}{[(\omega_0^2 - \omega^2)^2 + (\frac{\omega_0 \omega}{Q})]^2} \qquad (3)$$

Fig. 1(c) shows the $Q$ (blue squares) and corresponding gain ($Q_p/Q_o$) (red triangles) at different $V_p$. A clear $Q$ enhancement was observed as $V_p$ gradually increases. The maximum enhancement of $Q$ was achieved at $V_p = 25$ mV with $Q=235±9$ (blue curve), showing remarkably a 10-fold enhancement compared to the signal without pumping (red curve, $Q=24±1$). We further compare our $Q$ value with previous works on doubly-clamped SWNT resonators[8-10,20-23] in Fig. 1(d). Overall, previous room temperature $Q$ record is around 200, while our highest $Q$ through parametric amplification is ~700 (marked as a star and the resonance signal with fitting curve is shown in inset), showing at least three times improvement.

Next, we examined effects of system parameters on parametric amplification by looking at the DC gate voltage dependence and the AC driving voltage dependence on parametric amplification. Figure 2 (a) and 2(c) show the maximum gains at different $V_g$ obtained from two SWNT resonators, respectively. On both devices, we consistently observed effective parametric amplification with gain between 2 to 4 at higher $V_g$ (blue triangles), but no amplification with gain ~ 1 at lower $V_g$ (red triangles). To understand this disruption of parametric amplification, we plotted resonant frequency vs. $V_g$ for both resonators in the Figure 2(b) and 2(d), respectively. The resonance frequency is up-shifted at higher potential due to elastic hardening[18,20], and two vibrational regimes, bending and catenary regimes, are clearly observed[20]. Comparing the gain obtained at different $V_g$ with the corresponding vibration regimes, we found that efficient parametric amplification only occurs in the catenary regimes for both resonators. This can be explained by the much stronger spring constant modulation, $\frac{dk}{dV_g} \propto f \frac{df}{dV_g}$, in catenary regime. Extracting $df/dV_g$ from Fig. 2(b) and 2(d) suggests 3~8 times greater tunability in catenary regime than in bending regime. It is possible to achieve effective parametric amplification in the bending regime by applying large pumping voltage, in which case nonlinear effects need to be considered.

We also examined the effect of excitation driving voltage ($V_{sd}$) on parametric amplification, and the results of maximum gain vs. $V_p$ with different $V_{sd}$ at a fixed DC gate voltage ($V_g$= -4 V) are plotted in Fig. 2(e). The experimental data did not reveal any dependence between maximum gain and $V_{sd}$ when the $V_{sd}$ was increased from 20 mV to 60 mV. This result is consistent with observations from previous work, where no dependence was observed between the gain and $V_{sd}$[13].

Last, we examine the threshold voltage for parametric amplification ($V_t$). As shown in Figure 3(a), gain increases sharply with increasing $V_p$, agreeing with optimum parametric amplification near the threshold voltage. In order to extract $V_t$ from the measurement data, we note that equation (2) is derived under fixed phase lag between driving and pumping. However, our two-source FM technique will introduce a time varying phase lag, and hence the overall gain is an average result due to varying phases. Therefore, the average gain can be written as

$$G = \frac{1}{2\pi}\int_0^{2\pi} G(\phi)d\phi = \frac{1}{2\pi}\int_0^{2\pi}\left[\frac{\cos^2\phi}{(1+V_p/V_t)^2} + \frac{\sin^2\phi}{(1-V_p/V_t)^2}\right]^{1/2} d\phi \qquad (4)$$

Fitting data in Fig. 3(a) with equation (4) (red curve), we find a $V_t = 0.24$ V at $V_g = -5$ V. Furthermore, the average gain under constant pumping frequency at $2\omega_0$ and varying driving frequency can be calculated using $G = (1-(V_p/V_t)^2)^{-1}$ from ref [15]. The analytical fitting (green curve) yields $V_t = 0.23$ V at $V_g = -5$ V, consistent with the numerical fitting result. We further extract $V_t$ values under various $V_g$, and plot them in Fig. 3(b) (red squares). To model $V_t$ dependence on $V_g$, we follow the derivation of Rugar and Grütter[12] and have $V_t$ expressed as $V_t = 2k_0/QV_gC_g''$, where $C_g''$ is the second derivative of the gate capacitance with respect to the distance between the nanotube and gate. To calculate the theoretical values of $V_t$, we adopt the cylinder over an infinite plane model for capacitance, $C = \frac{2\pi\varepsilon_0 L}{\ln(\frac{2Z}{d})}$, and the results are plotted as blue diamonds in Figure 3(b). The experimental results show good agreement with the theoretical $V_t$ values, suggesting the possibility to predict and control $V_t$ by changing the device geometries and gate coupling. We also notice the deviation from the theoretical $V_t$ at $V_g = -6$ V, and its origin is not clear at this time.

Our results enable the light-weight carbon nanotube as high-Q NEMS resonator for single molecule and atomic mass sensing. We also expect the parametric amplification technique can be applied to other low-Q NEMS resonators suffering from intrinsic loss mechanisms, such as graphene resonators. The $2\omega_0$ modulation through electrostatic gating offers a simple technique which can be easily adopted in various device geometries and the flexibility to be integrated with NEMS applications.

Note Added:

Upon the completion of the manuscript, we became aware of a related work published online on Nano Letters.[24] Both works achieve parametric amplification by applying pumping voltages on the gate electrode, but with different measurement schemes and different focuses.

**Acknowledgements**

The work is supported by the start-up fund provide by the University of Michigan. This work used the Lurie Nanofabrication Facility at University of Michigan, a member of the National Nanotechnology Infrastructure Network funded by the National Science Foundation.

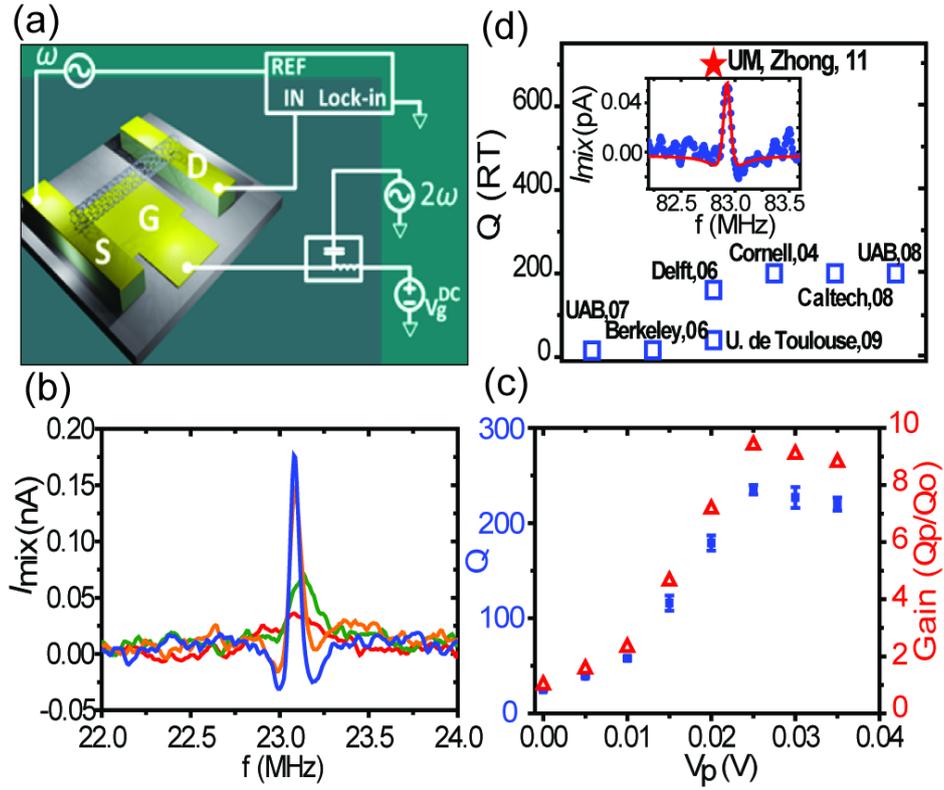

FIG. 1. (Color online). (a) Schematic diagram of the experimental setup for parametric amplification in SWNT NEMS resonators. (b) Frequency modulated mixing currents are plotted as a function of driving frequency at different pumping voltages ranging from 0 (red), 10 (green), 20 (orange), to 25 (blue) mV. (c) $Q$ (blue squares) and gain ($Q_p/Q_o$) (red triangles) vs. $V_p$. (d) List of maximum $Q$'s (blue squares) reported at room temperature in previous literatures. Our maximum Q achieved through parametric amplification is around 700 (marked as star) and the resonance signal (blue dots) with fitting curve (red) is shown in the inset.

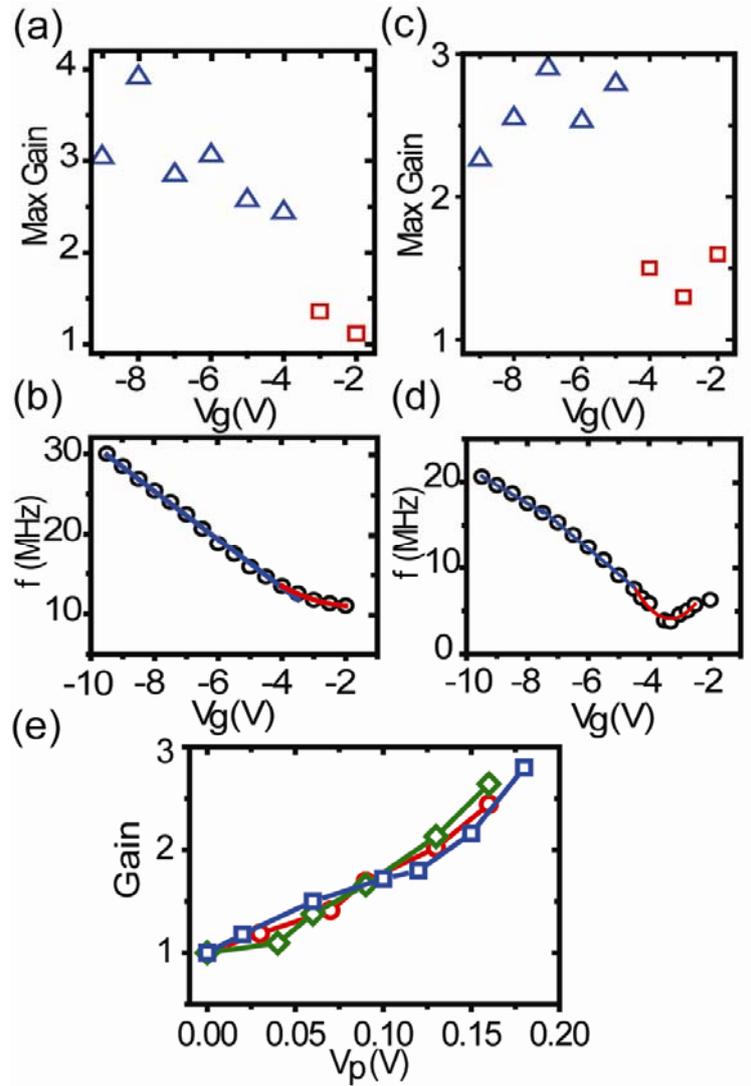

FIG. 2. (Color online). (a), (c) Maximum gain vs. $V_g$ of two SWNT resonators. (b), (d) The resonant frequency vs. $V_g$ for two SWNT resonators. Two vibration modes, bending (red) and catenary (blue) regimes, are shown clearly in both devices. (e) Gain vs. $V_p$ at different driving voltages ($V_{sd}$) ranging from 20 (red), 40 (green) to 60 (blue) mV.

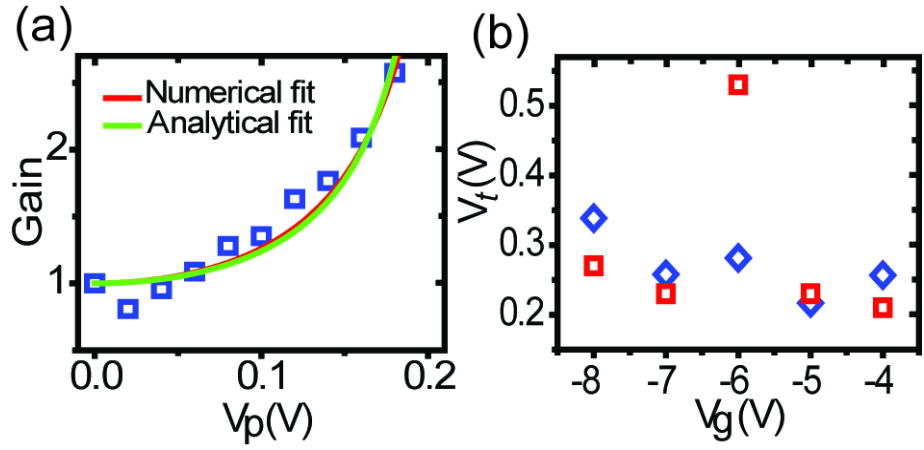

FIG. 3. (Color online). (a) Gain (blue squares) vs. $V_p$ at $V_g$= -5 V. The green curve is the analytical fit and the red curve is numerical fit by using equation (4). (b) Comparison of calculated $V_t$ (blue diamonds) and measured $V_t$ (red squares) at different $V_g$.